# A Two-stream End-to-End Deep Learning Network for Recognizing Atypical Visual Attention in Autism Spectrum Disorder

Jin Xie[1], Longfei Wang[1], Paula Webster[2], Yang Yao[1], Jiayao Sun[1], Shuo Wang[2*] and Huihui Zhou[1*]

[1] Brain Cognition and Brain Disease Institute, and Shenzhen-Hong Kong Institute of Brain Science, Shenzhen Institutes of Advanced Technology, Chinese Academy of Sciences.

[2] Department of Chemical and Biomedical Engineering, and Rockefeller Neuroscience Institute, West Virginia University.

\* Correspondence: Shuo Wang (wangshuo45@gmail.com); Huihui Zhou (hh.zhou@siat.ac.cn)

**Abstract**
Eye movements have been widely investigated to study the atypical visual attention in Autism Spectrum Disorder (ASD). The majority of these studies have been focused on limited eye movement features by statistical comparisons between ASD and Typically Developing (TD) groups, which make it difficult to accurately separate ASD from TD at the individual level. The deep learning technology has been highly successful in overcoming this issue by automatically extracting features important for classification through a data-driven learning process. However, there is still a lack of end-to-end deep learning framework for recognition of abnormal attention in ASD. In this study, we developed a novel two-stream deep learning network for this recognition based on 700 images and corresponding eye movement patterns of ASD and TD, and obtained an accuracy of 0.95, which was higher than the previous state-of-the-art. We next characterized contributions to the classification at the single image level and non-linearly integration of this single image level information during the classification. Moreover, we identified a group of pixel-level visual features within these images with greater impacts on the classification. Together, this two-stream deep learning network provides us a novel and powerful tool to recognize and understand abnormal visual attention in ASD.

## 1. Introduction

Eye movement studies have been common in the field of research on abnormal attention behaviors in people with Autism Spectrum Disorder (ASD) [1,2]. The reduced attention to socially relevant stimuli [3-8], the increased image center bias [9], and the impaired joint attention [10-15] have been reported in people with ASD. For example, Wang et al. [16] showed that ASD people had a stronger attention bias toward image center regardless of object distribution and reduced attention to face stimuli. However, these atypical attention behaviors were revealed by statistical comparisons of eye movements between ASD and Typically Developing (TD) groups. Because of the large variability of eye movements within ASD and TD groups, overlaps in eye movement behaviors across the groups, and inconsistency in some evidence for abnormal eye movements with ASD [17], it is difficult to accurately separate ASD from TD at the individual level based on these limited eye movement features [18].

Recently, the deep learning technology such as Convolutional Neural Network (CNN) has been highly successful in solving complex classification problems [19-24], which automatically extracts features for classification through a data-driven learning process without manually designing or selecting features. Machine learning has been applied to discern between those with ASD and TD based on brain activities, behavioral features such as motor behaviors, facial expressions, body gestures, and eye movements [25-31]. Two studies recognized ASD using eye-movement data [32,33], achieving 0.89 and 0.85 classification accuracy, respectively. Liu et al. [32] applied a traditional machine learning framework of K-means + Support Vector Machine (SVM) to identify those with ASD based on eye movements while observing single face stimuli. Jiang et al. [33] designed a CNN + SVM framework, in which the CNN was trained to reconstruct the eye movement pattern difference between ASD and TD groups, and the SVM used these pattern difference features from the CNN and eye movement features of a subject for classification. During testing, the inputs to the SVM included data from the test subjects and the subjects used for training, which is different from typical classifications that are based only on the test data. Therefore, there is still a lack of end-to-end deep learning framework for the ASD recognition based on eye movement data.

In this study, we developed a novel two-stream deep learning network using 700 natural scene images and corresponding Human Fixation Maps (HFMs) in order to discern between those with ASD and TD. By leveraging both image and eye movement information, the cropping and flipping data augmentation and the transfer learning technology, the classification accuracy of our model achieve 0.95, which is higher than accuracies in the previous state-of-the-art studies using eye movement data. Further, we estimated contribution of each natural scene image and corresponding HFM to the classification and integration of information from multiple images in our model. This study provides a novel high accuracy deep learning method for ASD research.

## 2. Methods

**2.1 Subjects**

Twenty high-functioning subjects with ASD and 19 TD neurologically and psychiatrically healthy subjects were recruited, and the two groups were matched on age, IQ, gender, race, and education (see [16] for details). Autism was evaluated using the ADOS (Autism Diagnostic Observation Schedule) [34] and the ADI-R (Autism Diagnostic Interview-Revised) [35]. All ASD participants met the DSM-IV/ICD-10 diagnostic criteria.

**2.2 Task and dataset**

In the eye-tracking experiment, subjects freely viewed 700 natural scene images from the OSIE database [36] for 3 seconds each image in random orders. Eye movement data was collected with a non-invasive infra-red remote Tobii X300 system at a sample rate of 300 Hz while participants were viewing these images. The pixel dimension of images is 800 × 600 pixels. Based on eye movement data from a subject observing an image, we constructed an 800 × 600 Human Fixation Map (HFM) for containing the total gaze time at each pixel location. Then, the HFM was smoothed with a Gaussian filter and normalized to the range from 0 to 1 [16]. We calculated HFMs for all subjects observing these 700 images. Fig. 1 showed HFMs of all subjects during observing a natural scene image which was plotted in the lower right corner of Fig. 1a. There was significant within-group variability in eye movement patterns for both groups, and substantial overlapping of the patterns across the two groups. The HFMs were calculated in the Matlab platform.

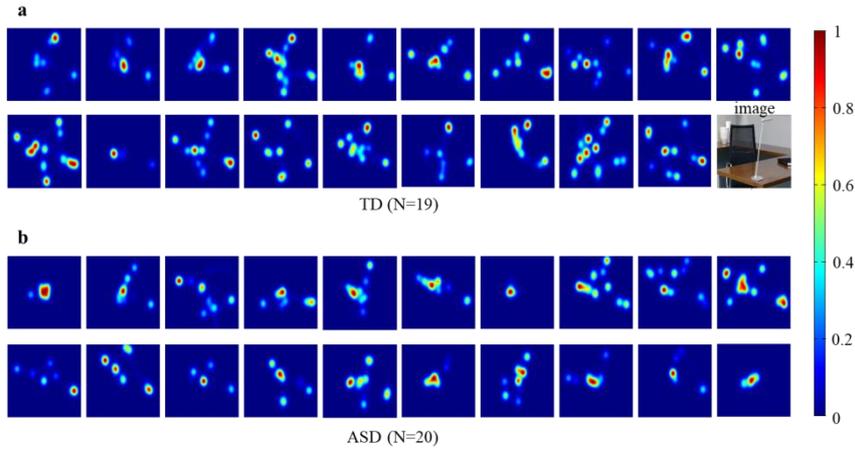

**Fig. 1 | HFMs of all subjects during observing a natural scene image. a**, TD subjects. **b**, ASD subjects.

**2.3 Two-stream deep learning network for ASD recognition**

Our two-stream network (as shown in Fig. 2) included two same VGGNets modified from VGG16 [37], one of commonly used CNN architectures for object recognition. We removed the last fully connected layer from VGG16, therefore, the VGGNet included 13 convolutional layers (Conv), 5 Max-pooling layers and 2 fully connected layers (Fc), which generated a 700 × 4096 feature map as its output. The two feature maps from the two VGGNets were combined and feed to a three-layer network (ASDNet) including 1 convolutional layer and 2 fully connected layers. Together, our two-stream network had 18 layers except for the Max-pooling layers. At the end of our two-stream network, a softmax layer was applied to transform real values into probabilities. The images and corresponding HFMs of subjects were feed into the network simultaneously. In the two-stream network, eye movement features and image features were extracted by the two VGGNets, respectively, and latter combined for ASD recognition.

We adapted a transfer learning strategy by using the VGGNet [37] pre-trained on the ImageNet challenge dataset [38]. During training, we fixed weights in the VGGNet, and trained the ASDNet from scratch with back propagation to optimize the network for ASD recognition. In the ASDNet, the convolutional layer had a 1 × 1 filter, and the length of two fully connected layers were 512 and 2, respectively. The activation function of rectified linear unit (ReLU) [19] was used in the two-stream network. Our two-stream network was built on the Caffe platform [39].

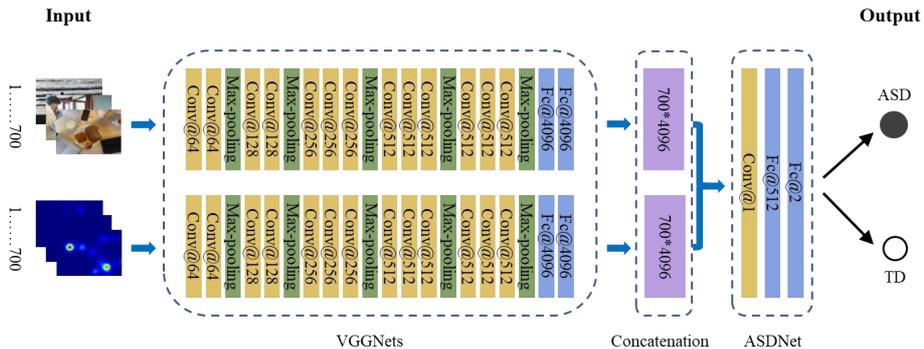

**Fig. 2 | The architecture of the two-stream network for ASD recognition.**

**2.4 Data Augmentation**

To improve the performance of our two-stream network, we applied the data augmentation by cropping and flipping the original dataset with label preserved [19]. The pre-trained VGGNet required a fixed-size input of 224 × 224. The original size of the natural scene images and HFMs were 800 × 600, which were resized to 256 × 256. We extracted five 224 × 224 crops from each 256 × 256 image covering its 4 corners and its center, respectively [40]. Flipping augmentation were conducted by mirroring all crops across their vertical axes. Together, we obtained 10-fold augmentation of the original dataset. We applied the same augmentation procedures in our training and test data. The data augmentation was implemented in the Matlab platform.

**2.5 t-distributed Stochastic Neighbor Embedding (t-SNE) visualization**

To visualize internal features learned by the two-stream CNN, we applied a t-SNE method to convert high-dimensional features in the two-stream network into 2-dimensional maps. t-SNE is a variation of Stochastic Neighbor Embedding (SNE) [41], a commonly used method for multiple class high-dimensional data visualization [42]. We applied t-SNE visualization for the last two fully connected layers of our two-stream network, converting the 512-dimensional and 2-dimensional representations in the two layers to two 2-dimensional maps, respectively. The two maps were scatterplots with one point representing one subject. We implemented the t-SNE in the Matlab platform.

**2.6 Layer-Wise Relevance Propagation (LRP)**

We applied a Layer-Wise Relevance Propagation (LRP) method [43] to visualize and identify pixel-wise information important for ASD recognition by our network, which revealed the contribution of each pixel in an image to the classifier prediction. In our analysis, the output of the last fully connected layer was reversely transmitted through ASDNet and VGGNets to our network inputs to obtain LRP maps for 700 images and HFMs. Each LRP map was represented by a conventional heat-map. We focused on the LRP maps for the images to identify visual features important for ASD recognition. The pixels in an image were important for recognition if their absolute LRP values were higher than a threshold, which was 0.0029. On average, the LRP summation of these important pixels was 75% of the LRP summation of all pixels in the images. Therefore, contribution from these important pixels consisted of 75% contribution of whole image on average. In the LRP map, we retained LRP values in these important areas and set values in other areas to zero. The LRP map was overlaid with its corresponding image to display important image features for ASD recognition. We implemented the LRP in the TensorFlow platform.

**2.7 Performance metric**

Because of the 10-fold augmentation, each subject was tested 10 times. The trained model correctly recognized a subject when classification score of the subject was $\geq 0.6$, and the classification score was the proportion that the subject was correctly predicted in the 10 times. A test correctly predicted the subject if the probability of the subject belonging to its true group was $> 0.5$. The model-level accuracy was the averaged classification score across all test subjects in the leave-one-out or 13-fold cross-validation (see Experiment 3.3). The subject-level accuracy (Acc.) was the ratio of correctly recognized test subjects to all test subjects in the leave-one-out or 13-fold cross-validation. We also calculated the Sensitivity (Sen.) and Specificity (Spe.) according to [44]. The sensitivity measured the rate of ASD subjects correctly predicted as ASD, while the specificity measured the rate of TD subjects correctly predicted as TD. We plotted the Receiver Operator Characteristic (ROC) curve by introducing different probability thresholds, and computed the corresponding Area Under the Curve (AUC) [45] to evaluate classification performance of our two-stream network.

The AUC was also used to measure the contribution of single image data (image + corresponding HFM) to network prediction, and the contribution of multiple image data. In our two-stream network, after the 700 images and corresponding HFMs passing through the VGGNet model, we obtained two 700 × 4096 feature maps: one for 700 images and one for 700 HFMs. To estimate the contribution of a single image data to classification, we only retained the two 1 × 4096 feature vectors representing the image and corresponding HFM, and set the other 699×4096 feature vectors in the two 700 × 4096 feature maps to zero to calculated ASD probability of a subject. The AUC of this single image data was calculated based on the ASD probabilities of all 39 subjects. 700 AUCs were calculated for all 700 single image data. A baseline AUC was calculated when the two 700 × 4096 feature maps were set to zero. The contribution of a single image data was positive when its AUC was greater than the baseline AUC. Similarly, we evaluated the contribution of multiple image data by retaining multiple 1 × 4096 feature vectors in two 700 × 4096 feature maps. We used the sklearn library in Python to calculate the AUC.

## 3. Experiment

**3.1 Environment**

All the works were conducted on a multi-GPU server equipped with two 16-core Intel Xeon E5-2650 V4 processors, 256 GB memory, and nine NVIDIA GeForce GTX 1080 Ti with 99 GB memory totally. The server operated in the Ubuntu16.04 operating system.

**3.2 Training**

Our two-stream network ran on the deep learning framework of Caffe [39]. During the training, we fixed the weights of pre-trained VGGNet and trained the ASDNet in our two-stream network with back-propagation [46]. Training was performed by the Stochastic Gradient Descent (SGD) optimizer with the base learning rate of $10^{-5}$ and the learning policy of "inv" [39]. The weight initializers of the ASDNet were "xavier" and "gaussian" [39]. 50% dropout was applied to avoid overfitting. The maximum number of training iterations was 1000.

### 3.3 Cross-Validation

Leave-one-out and 13-fold cross-validation were used in this study. During the leave-one-out cross-validation, one out of 39 subjects was selected sequentially for testing, and the model was trained on the dataset from the remaining 38 subjects, which returned the probabilities to be ASD and TD of the test subject. There was no overlap between training and test dataset. Such a process repeated 39 rounds. During the 13-fold cross-validation, the 39 subjects were randomly split into 13 3-subject subsets. One of the 13 subsets was selected sequentially for testing, and the model was trained on the dataset from the 36 subjects of the remaining 12 subsets, which returned the probabilities to be ASD and TD of these three test subjects. We repeated this process for 13 rounds.

## 4. Results

### 4.1 Classification performance

The model-level accuracy and cross-entropy loss were plotted against the training iterations in the leave-one-out and 13-fold cross-validation (Fig. 3). The classification accuracies on training and test datasets gradually increased to stable levels, whereas the corresponding losses gradually decreased to stable levels as the training iterations increasing. Our two-stream network achieved 0.92 and 0.84 accuracy in the leave-one-out and 13-fold cross-validation tests, respectively. We plotted the ROC curves in the leave-one-out and 13-fold cross-validation after 10-fold data augmentation (Fig. 4). In the leave-one-out cross-validation, our model obtained 0.95 subject-level accuracy (Acc.), 1.00 sensitivity (Sen.), 0.89 specificity (Spe.) and 0.93 AUC, respectively. In the 13-fold cross-validation, the subject-level accuracy (Acc.) reached 0.85, with 0.80 sensitivity (Sen.), 0.89 specificity (Spe.) and 0.91 AUC, respectively.

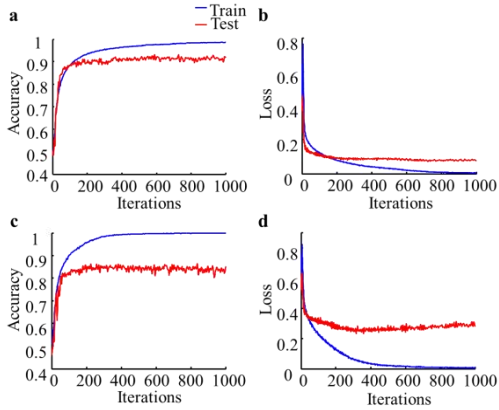

**Fig. 3 | The accuracy and loss curves of our network.**
**a-b**, In the leave-one-out cross-validation.
**c-d**, In the 13-fold cross-validation.

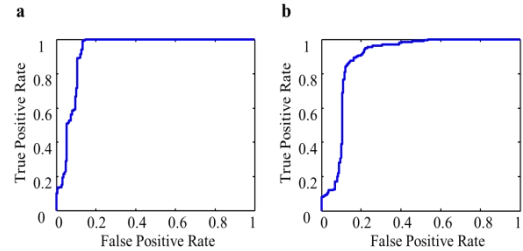

**Fig. 4 | The ROC curves of our two-stream network.**
**a**, In the leave-one-out cross-validation.
**b**, In the 13-fold cross-validation.

We compared our method with the state-of-the-art [32,33]. Liu et al. [32] used a K-means + SVM method and Jiang et al. [33] adopted a CNN + SVM framework to recognize ASD using eye movement data. In this study, we developed a new end-to-end two-stream deep learning network. Overall, our method outperformed the state-of-the-art in all performance metrics (Table. 1).

**Table. 1 | A quantitative comparison between our model and the previous state-of-the-art models.**

| Methods | Models | Cross-validation | Acc. | Sen. | Spe. | AUC |
| --- | --- | --- | --- | --- | --- | --- |
| Liu et al. [32] | K-means + SVM | leave-one-out | 0.89 | 0.93 | 0.86 | 0.90 |
| Jiang et al. [33] | CNN + SVM | leave-one-out | 0.85 | 0.83 | 0.87 | 0.89 |
| Ours | Two-stream network | leave-one-out | 0.95 | 1.00 | 0.89 | 0.93 |
| | | 13-fold (three-out) | 0.85 | 0.80 | 0.89 | 0.91 |

### 4.2 t-SNE visualization

To examine the internal features learned by the two-stream network, we visualized the high-dimensional features through t-SNE [42] by a two-dimensional scatter plot with each point representing a subject (Fig. 5). The t-SNE was applied to the last two layers of ASDNet with 512 and 2 dimensions of features, respectively. Fig. 5a and Fig. 5b showed a typical example in the leave-one-out and 13-fold cross-validation, respectively. In the leave-one-out cross-validation, t-SNE distribution of data from ASD and TD subjects was well separated in the last two layers, demonstrating that the features from two-stream network were robust and unambiguous. In the 13-fold cross-validation, the vast majority of data from ASD and TD subjects were well separated, with one subject misclassified. Fig. 5 showed that these features learned by the leave-one-out model were more effective than those learned by the 13-fold model. In addition, better separation was found in the last layer, suggesting that ASD and TD features became more distinguishable along the network.

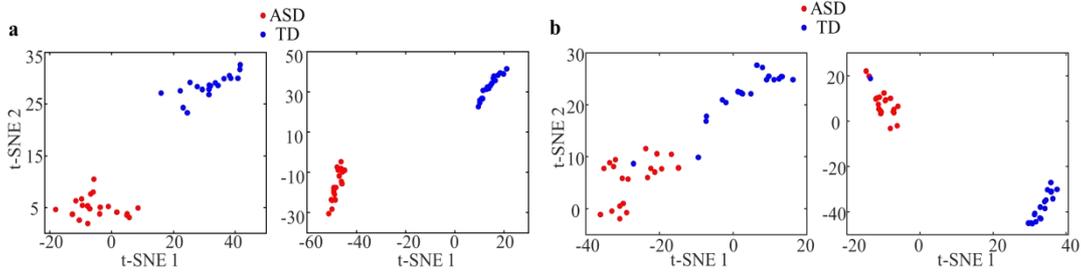

**Fig. 5 | t-SNE visualization of high-dimensional features in our two-stream network. a**, One typical example in a leave-one-out model. Left: in the second last layer; Right: in the last layer. **b**, One typical example in a 13-fold model. Left: in the second last layer; Right: in the last layer.

**4.3 Contribution of single image data to the classification**

The previous classification results relied on the 700 images and corresponding HFMs. We analyzed the contribution of a single image data (a natural scene image and its corresponding HFM) by retaining feature vectors representing the single image data, and setting the other feature vectors representing the remaining 699 natural scene images and HFMs to zero, and an AUC measurement was used to evaluate this contribution (see details in Methods 2.7). We plotted the AUC values of all 700 image data in a descending order in Fig.6a. The black dotted line represented baseline level of the AUC value, which was calculated when the feature maps representing the 700 natural scene images and HFMs were set to zero. The contribution of a single image data was positive, when its AUC value was greater than the baseline value, otherwise its contribution was negative. More than 500 of 700 single image data had a positive contribution, which seemed consistent with the high classification accuracy of our two-stream network. However, the AUC values from the single image data ( < 0.55) were still far smaller than the AUC value of our two-stream network (0.93) using all 700 natural scene images and HFMs, suggesting combination of multiple single image data played an important role in classification. Fig. 6b showed these natural scene images of AUC top-10 and bottom-5.

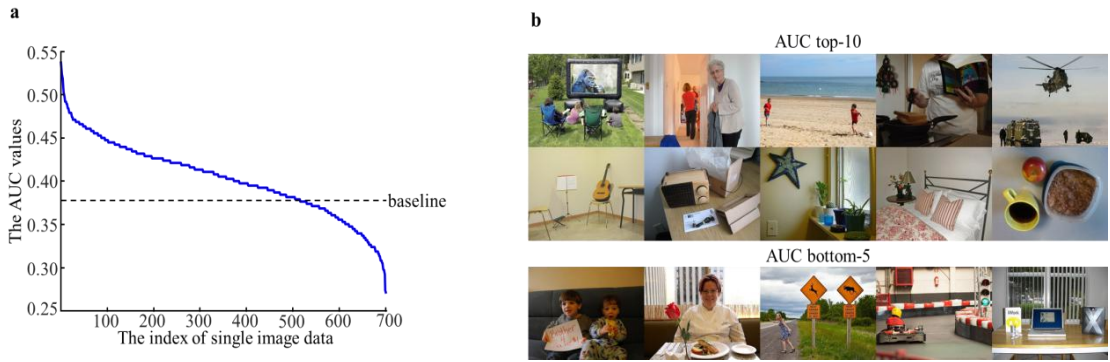

**Fig. 6 | Contribution result of single image data to the classification. a**, The AUC values of single image data: the horizontal axis represented the index of each single image data in this descend order of AUC values, and the vertical axis represented the AUC values of the single image data. **b**, These natural scene images of AUC top-10 and bottom-5.

**4.4 Classification performance based on multiple image data**

Based on the previous sorted AUC values of single image data, we calculated AUCs of multiple image data including different numbers of top single image data according to its AUC value (Fig. 7a). The AUC was about 0.71 based on the top-50 image data. With the increase of the number of single image data included in the multiple image data, AUC gradually increased to 0.95 from top-150, and reached the peak value of 0.99 from top-250, and stayed near to the peak value from top-300 and top-350 image data. Further addition of image data did not improve the performance. In fact, the AUC value from data of all 700 images was 0.93, which was smaller than the peak value from the top-250 image data. This result indicated that the combined information of 700 image data was redundant, and information from multiple image data was not linearly integrated from each single image data, but with both coordination and antagonism interactions between them. We used a method of step by step discarding single image data in iterations from top-150, and found a combination of 16 single image data with an AUC of 0.87. The 16 natural scene images were shown in Fig. 7b.

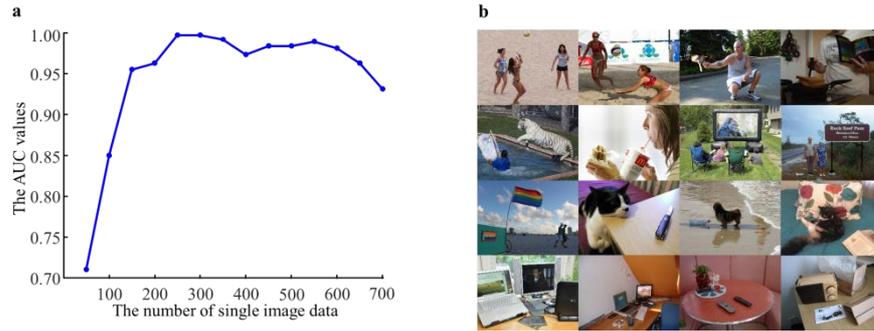

**Fig. 7 | Classification result based on multiple image data. a**, The AUCs based on different multiple image data: the horizontal axis represented the number of single image data included in the multiple image data, and the vertical axis represented the AUC values of multiple image data. **b**, Natural scene images: the AUC based on these images and corresponding HFMs was 0.87.

**4.5 Contribution of the pixel-level visual features within the natural scene images to the classification**

We applied the LRP method to visualize the pixel-level visual features within the natural scene images important for classification by our two-stream network. We selected 100 natural images with highest LRP value summation from the top-300 single image data (see Fig. 6). From LRP maps of these 100 images, we identified 11 types of image features often containing pixels important for classification (as indicated by red and blue points in Fig. 8) with LRP values higher than the threshold (see Methods 2.6). These 11 types of image features were: human faces (Fig. 8 a-e), human arms and hands (Fig. 8 a-e), upper bodies (the human body of below the neck and above the waist; Fig. 8 a, c, e), lower bodies (the human body of below the waist and above the knee; Fig. 8 b, c, e), the lower part of limbs and foots (the human body of below the knee; Fig. 8 b, c, e, f), objects at attention focus (what the person in the image is looking at; Fig. 8 b, c, d), animal faces (Fig. 8 e, g-i), animal bodies(Fig. 8 e, g, h), cups (Fig. 8 j-k), fruits (Fig. 8 i-j), foods (Fig. 8 j-k). We calculated the LRP score for each type of image feature. For each appearance of this type of feature in the 100 natural scene images, we set a score between 0 and 1 based on the number of pixels important for classification contained by the feature, with the higher score associated with more pixels. We showed some examples with different LRP scores of the human faces feature in Fig. 9a. As shown in Fig. 9b, the averaged LRP scores of all 11 types of image features were significantly higher than the LRP score of uniform background (P < 0.05, Ranksum test).

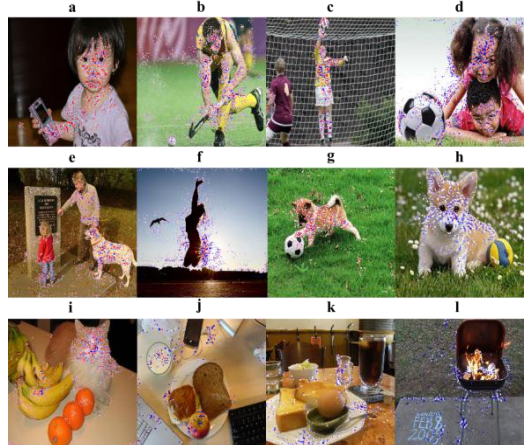

**Fig. 8 | Some examples of LRP maps visualized image features with important for classification.**

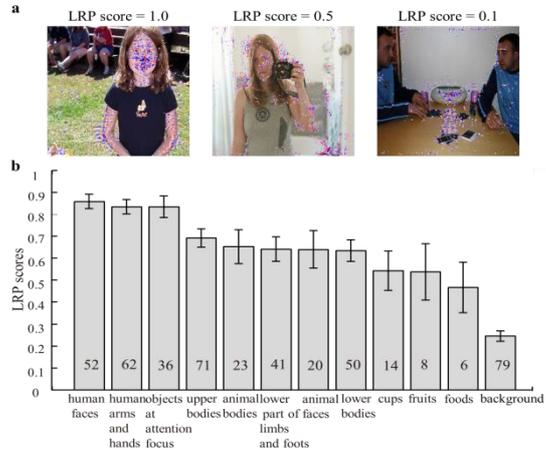

**Fig. 9 | The result of LRP scores. a**, Some examples with different LRP scores within human faces feature. **b**, The LRP scores of 11 types of image features.

## 5. Discussion

In this study, we developed a novel two-stream end-to-end deep learning network combining image and eye movement information for discerning between those with ASD and TD, obtaining accuracies of 0.95 and 0.85 in the leave-one-out and 13-fold cross-validation, respectively. We observed a well separation of high-level features representing ASD and TD subjects from the last two layers of our two-stream network by a t-SNE visualization method, suggesting that our model represented the features of ASD and TD data well after learning. Deep learning networks have been applied for ASD recognition with most studies based on fMRI signals using auto encoders with 3 to 8 layers [22,47-50], one or multi-stream CNN [31,51] etc, the accuracy of our model is comparable with the best accuracy of these deep learning models. Recently, Tao and Shyu [52] developed a SP-ASDNet including multi-stream CNNs and LSTM for classification between ASD and TD based on eye movement scanpath data, which obtained a 0.74 validation and 0.58 test accuracy that is lower than the performance of our model and Jiang et al.'s [33] model.

We characterized the contributions of single image data to the classification, which suggested that multiple image data was necessary for high accuracy recognition because of the small AUC values based on these single image data. We found the AUC of our model reached 0.87 based the data including 16 natural scene images and HFMs, suggesting that it was possible to recognize abnormal eye movements of ASD based a small but very informative data. Through LRP visualization, we identified pixel-level visual features in the natural scene images with greater impact on the classification including faces, different parts of human body, objects at attention focus, and some non-social features. So far, the integrating process of information from multiple image data and information from different parts of images are still not clear. Further study is needed to understand this information integration process, and design more informative and smaller image sets for recognizing atypical visual attention in ASD with higher efficiency.